\documentstyle[prl,aps,twocolumn]{revtex}

\draft

\input{psfig}

\begin{document}

\twocolumn[
\hsize\textwidth\columnwidth\hsize\csname @twocolumnfalse\endcsname

\title{
The one-pion-exchange three-nucleon force and the $A_y$ puzzle}

\author{  L. Canton$^{1,2}$ and W. Schadow$^3$ }
\address{$^1$Istituto Nazionale di Fisica Nucleare, Sez. di Padova,
             Via F. Marzolo 8, Padova I-35131, Italy \\
         $^2$Physics Department, University of Manitoba, Winnipeg, MB,
             Canada R3T 2N2 \\
         $^3$TRIUMF, 4004 Wesbrook Mall, Vancouver, British Columbia,
             Canada V6T 2A3}

\date{March 30, 2001}

\maketitle

\begin{abstract}
We consider a new three-nucleon force generated by the exchange of
one pion in the presence of a $2N$ correlation. The underlying irreducible
diagram has been recently suggested by the authors as a possible candidate
to explain the puzzle of the vector analyzing powers $A_y$ and $iT_{11}$
for nucleon-deuteron scattering. Herein, we have calculated the elastic
neutron-deuteron differential cross section, $A_y$, $iT_{11}$, $T_{20}$,
$T_{21}$, and $T_{22}$ below break-up threshold by accurately solving the
Alt-Grassberger-Sandhas equations with realistic interactions.
We have also studied how $A_y$ evolves below 30 MeV.
The results indicate that this new $3NF$ diagram provides one
possible additional contribution, with the correct spin-isospin
structure, for the explanation of the origin of this puzzle.
\end{abstract}

\vspace{2mm}

]

The $A_y$ puzzle (or, more appropriately, the puzzle of the
vector analyzing powers) is probably the most famous of the open
problems in $3N$ scattering at low energy. The problem with this observable
has been observed quite early at the Tokio \& Sendai Conference
(Few-Body XI, 1986) ~\cite{Koike86}, since at that time
the first reliable Faddeev calculations with realistic $2N$ potentials
were becoming available, thanks in particular to the employment
of separable expansion methods which transform the $3N$ scattering
equations of Alt, Grassberger, and Sandhas (AGS)~\cite{AGS} into an effective,
multichannel Lippmann-Schwinger equation. This method of calculation
has been pushed forward to obtain accurate results
particularly by the Graz group~\cite{Graz}.
Since then, various alternative methods of solution of the
$3N$ scattering equation have been developed and tested~\cite{Gloeckle96},
and outstanding progresses have been made in the computational
techniques in order to:
1) include three-nucleon forces ($3NF$) in the $3N$ scattering
equations~\cite{Gloeckle96,Witala94,Witala98};
2) treat explicitly the $\Delta$ dynamics in the $3N$ system~\cite{Nemoto98};
3) provide a combined description of the $3N$ dynamics with Coulomb,
realistic $2N$, and phenomenological $3N$ forces~\cite{Rosati96}.

The puzzle was confirmed by these new approaches, and it turned out
that the existing $2\pi$-$3NF$~\cite{TM,BR,UA} provided a too small
effect for $A_y$~\cite{Gloeckle96,Witala94,Rosati96}, and not always
in the right direction.  
In absence of new $3NF$'s that could explain the puzzle, and 
since the $3N$ $A_y$ is rather sensitive to the $^3P_j$ $NN$ phase shifts,
it was concluded in Ref.~\cite{Tornow98}
(see also references therein) that such phases and the associated
$NN$ potentials derived from modern phase-shift analysis must be modified
at low energy. 
These modifications can be achieved without affecting appreciably 
the $2N$ data because the low-energy $2N$ observables cannot 
resolve the $^3P_j$  phases uniquely due to the Fermi-Yang ambiguities.
However, as has been argued in Ref.~\cite{Huber98}, 
it is not possible to increase the $3N$ $A_y$ with reasonable 
changes in the $NN$ potential, hence additional 
$3NF$'s of new structure have to be considered.
Recently, an attempt has been
made~\cite{Kievsky} using a purely phenomenological $3NF$ of
spin-orbit type, constructed {\it ad hoc} to affect only the
triplet-odd states of $2N$ subsystem.  $3NF$ terms of
pion-range/short-range form\cite{Coon95} have been reconsidered
lately from the point of view of Chiral Perturbation
Theory~\cite{Huber99} ($\chi$PT), which predicts for these terms a
non-negligible role.  Qualitatively, it was found that these terms
somewhat affect $A_y$, but a quantitative conclusion could not be
derived.

It is our intention to indicate here a possible solution of this
puzzle in terms of a new $3NF$ proposed 
recently by the authors~\cite{CantonSchadow00}. This force is
generated by the one-pion-exchange diagram when one of the two
nucleons involved in the exchange process rescatters with a third one
while the pion is ``in flight''. The underlying diagram has been
derived starting from a formalism~\cite{Canton98} devoted to the
explicit treatment of the pion dynamics in the $3N$ system. The
resulting dynamical equation resembles a Faddeev-AGS equation, but
entails in its inner structure the full complexity of the underlying
four-body ($\pi NNN$) system.  A gradual procedure to project out the
pion degrees of freedom has been discussed in Ref.~\cite{CMS00}, where
it has been shown that this formalism leads to irreducible $3NF$
diagrams, and includes in particular the $3NF$ we will evaluate here.

$3NF$ diagrams of the type derived in Ref.~\cite{CantonSchadow00} and
analyzed here are not new in the
literature~\cite{Brueckner,Pask,Yang}, but they have been discarded in
modern few-nucleon calculations because of the presence of a
cancellation effect which has been observed in
Refs.~\cite{Yang86,Coon86} and discussed later from the point of 
view of effective chiral Lagrangians~\cite{vanKolck94}.
This cancellation involves meson retardation effects of the 
iterated Born term, and the irreducible diagrams generated by
sub-summing all time orderings involving the combined exchange of two
mesons amongst the three nucleons. However, this cancellation is
incomplete~\cite{CantonSchadow00} and generates a three-nucleon force
through a subtraction term of the type ``$t$-$v$'' with the 2$N$
$t$-matrix being pushed fully off-the energy shell because of the
presence of the pion.  Obviously, some ambiguities and model
dependencies should be expected, since the subtraction involves the
$t$-matrix in a region which is difficult to access and constrain,
e.g., by 2$N$ data.  In addition, there are also possible model
dependencies on how this ``imperfect'' cancellation should be
specified, since the input 2$N$ potential itself could possibly
include - maybe in somewhat hidden way - meson-retardation correction
effects. Ideally, this new $3NF$ term should be constructed
consistently with the specific aspects and details of the given 2$N$
interaction.

The explicit expression of this force~\cite{CantonSchadow00} is
\begin{eqnarray}
\label{OPE-$3NF$}
\lefteqn{V^{3N}_3({\bf p,q,p',q'};E) =
 {f_{\pi NN}^2(Q)\over m_\pi^2}{1\over (2\pi)^3}}
 \\ && \times
  \!\left [
{
({\mbox{\boldmath $\sigma_1$}}\cdot{\bf Q})
({\mbox{\boldmath $\sigma_3$}}\cdot{\bf Q})
({\mbox{\boldmath $\tau_1$}}\cdot {\mbox{\boldmath $\tau_3$}})
+
({\mbox{\boldmath $\sigma_2$}}\cdot{\bf Q})
({\mbox{\boldmath $\sigma_3$}}\cdot{\bf Q})
({\mbox{\boldmath $\tau_2$}}\cdot {\mbox{\boldmath $\tau_3$}})
\over
\omega_\pi^2
}
\right]  \nonumber \\
&&\times \, {
\tilde t_{12}({\bf p},{\bf p'};E-{{q}^2\over 2\nu} - m_\pi)
\over
2m_\pi}
\nonumber \\
&& +{f_{\pi NN}^2(Q)\over m_\pi^2}{1\over (2\pi)^3}
\, {
\tilde t_{12}({\bf p},{\bf p}';E-{{q'}^2\over 2\nu} - m_\pi)
\over
2m_\pi}
\nonumber   \\
& & \!\!
 \times \! \left [
{
({\mbox{\boldmath $\sigma_1$}}\cdot{\bf Q})
({\mbox{\boldmath $\sigma_3$}}\cdot{\bf Q})
({\mbox{\boldmath $\tau_1$}}\cdot {\mbox{\boldmath $\tau_3$}})
+
({\mbox{\boldmath $\sigma_2$}}\cdot{\bf Q})
({\mbox{\boldmath $\sigma_3$}}\cdot{\bf Q})
({\mbox{\boldmath $\tau_2$}}\cdot {\mbox{\boldmath $\tau_3$}})
\over
\omega_\pi^2
}
\right ] \, . \nonumber
\end{eqnarray}
The full $3NF$ is given by summing over
the cyclic permutations of
the nucleons,
$V^{3N}=V_1^{3N}+V_2^{3N}+V_3^{3N}$.
The momenta ${\bf p,q}$ represent respectively the Jacobi coordinates
of the pair ``$1$$2$'', and spectator ``$3$'', while $E$ is the 3$N$ energy.
We have set the pion-nucleon coupling constant to the traditional
value $f_{\pi NN}^2/(4\pi)$~=~0.078, and have employed
standard form-factors of monopole type to describe the
effective, composite nature of the meson-baryon coupling:
\begin{equation}
f_{\pi NN}(Q)=f_{\pi NN}{\Lambda_\pi^2-m_\pi^2 \over \Lambda_\pi^2 +Q^2} \, .
\end{equation}
The chosen $2N$ model interaction implicitly determines  
the value of pion-nucleon cut-off in $V^{3N}$, {\it e.g.} 
$\Lambda_\pi=1.7$ GeV for the Bonn {\em B} potential.
The transferred momentum ${\bf Q}={\bf q'}-{\bf q}$ enters
also in $\omega_\pi=\sqrt{m_\pi^2+Q^2}$.
$\tilde t_{ij}$ denotes the subtracted $t$-matrix between nucleons $1$ and $2$,
defined according to the prescription
\begin{equation}
\tilde t_{12}({\bf p},{\bf p'};Z)=c(Z)\, t_{12}({\bf p},{\bf p'};Z)
-v_{12}({\bf p},{\bf p'}) \, .
\label{SUB-1}
\end{equation}

Other details can be found in Ref.\cite{CantonSchadow00}.

In addition, we have introduced here the effective parameter $c(Z)$,
which is the only adjustable quantity of this $3NF$. This parameter
represents an overall correction factor for the far-off-the-energy-shell
$2N$ $t$-matrix entering this $3NF$ diagram. Ideally, $c(Z)$ 
should be one for a $2NF$ model able to provide a 
reliable extrapolation of the $t$-matrix down to $Z\approx -160$ MeV.
However, none 

\begin{figure}
\centerline{\hbox{
\psfig{figure=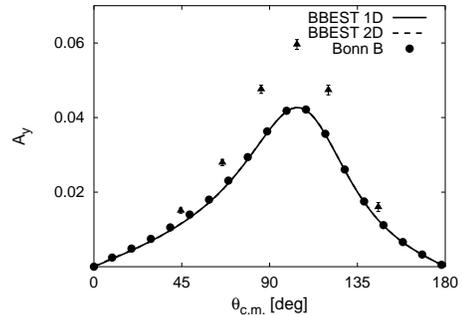,width=62mm,angle=-90}
}}
\vspace{2mm}
\caption[ ]{The $A_y$ puzzle, for
$nd$ scattering at 3 MeV (Lab).
Calculations with the Bonn {\em B} potential (dots), 
and with the high-rank BBEST potential (lines). 
With the BBEST potential there are two calculations, 
one obtained using the (non separable) 2-dim approach,
the other with the separable 1-dim algorithm. 
The curves are not distinguishable. Data (triangles) 
from Ref.~\cite{McAninch94}.}
\label{Ay_test}
\end{figure}
\noindent
of the existing $2N$ $t$-matrices can guarantee such
extrapolation since they are all constrained by experiments
at the deuteron pole and at $Z\ge 0$. Furthermore, $c(Z)$ might also
correct for possible model dependencies on how this imperfect
cancellation manifest itself, as already observed at the beginning of this
communication.
On general grounds, one expects that with increasing
energy in $nd$ scattering, the factor $c(Z)$ should drift towards one,
since the off-shell $2N$ $t$-matrix in $V^{3N}$ approaches gradually
the energy region with experimental constrains.

To calculate the $nd$ scattering observables below threshold,
we have used the high-rank BBEST potential
as two body input~\cite{Graz,Haidenbauerprivate}.
With this separable representations of the Bonn {\em B}
potential, it is possible to solve accurately the Faddeev-AGS
scattering equations, and obtain results comparable
(with errors less than 1\%) to those
obtained from a direct solution of the Faddeev equations using
the original potentials as input.
As an example, Fig.~\ref{Ay_test} shows the results we have
obtained for $A_y$ at 3 MeV with the BBEST potential and with
the original Bonn  interaction.
There are three curves since the  calculations  have  been 
performed using both the separable (1-dimensional) algorithm and the
non-separable (2-dimensional) method based on spline interpolation and
Pad\'e approximants, but the lines are practically
indistinguishable.  
Similar tests have been made also before~\cite{Cornelius90,Schadow98}
on various occasions.  Triangles represent the $nd$ experimental data
from Ref.~\cite{McAninch94}.

The three-nucleon
forces can be incorporated in the scattering equation
in a relatively simple way if the $2N$ input potential
is of finite rank.
We sketch the procedure for a rank one case:
once the separable $2N$ $t$-matrix is given,
$t=|g_1\rangle \tau \langle g_1|$,
and the (anti)symmetrized AGS equation have been rewritten in
the Lovelace form,
\begin{equation}
X_{11}=Z_{11}+Z_{11}\tau X_{11},
\end{equation}
the above $3NF$ is incorporated into this one-dimensional
\begin{figure}
\centerline{\hbox{
\psfig{figure=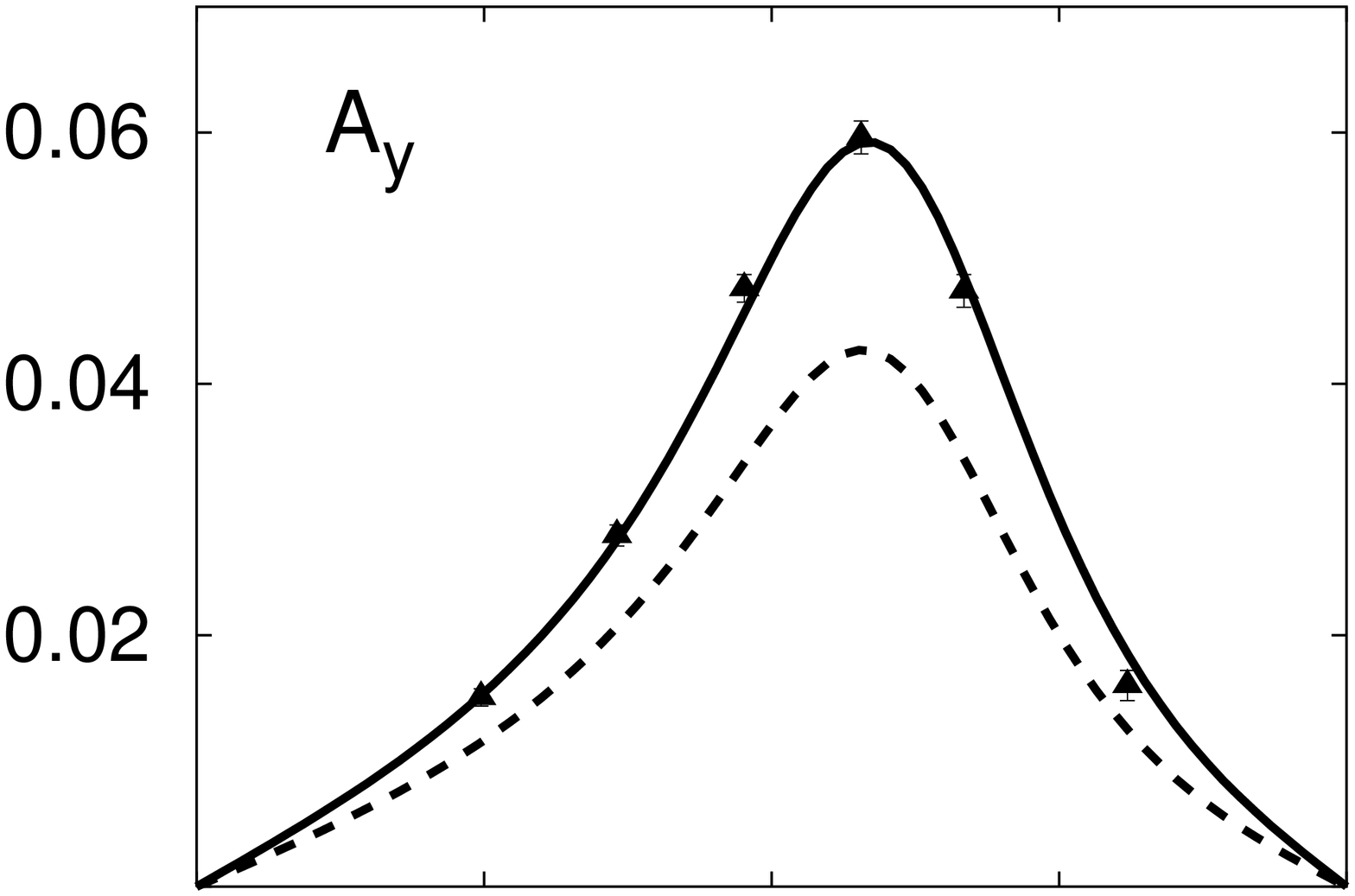,width=48mm,angle=-90}
\hspace{-1.0cm}
\psfig{figure=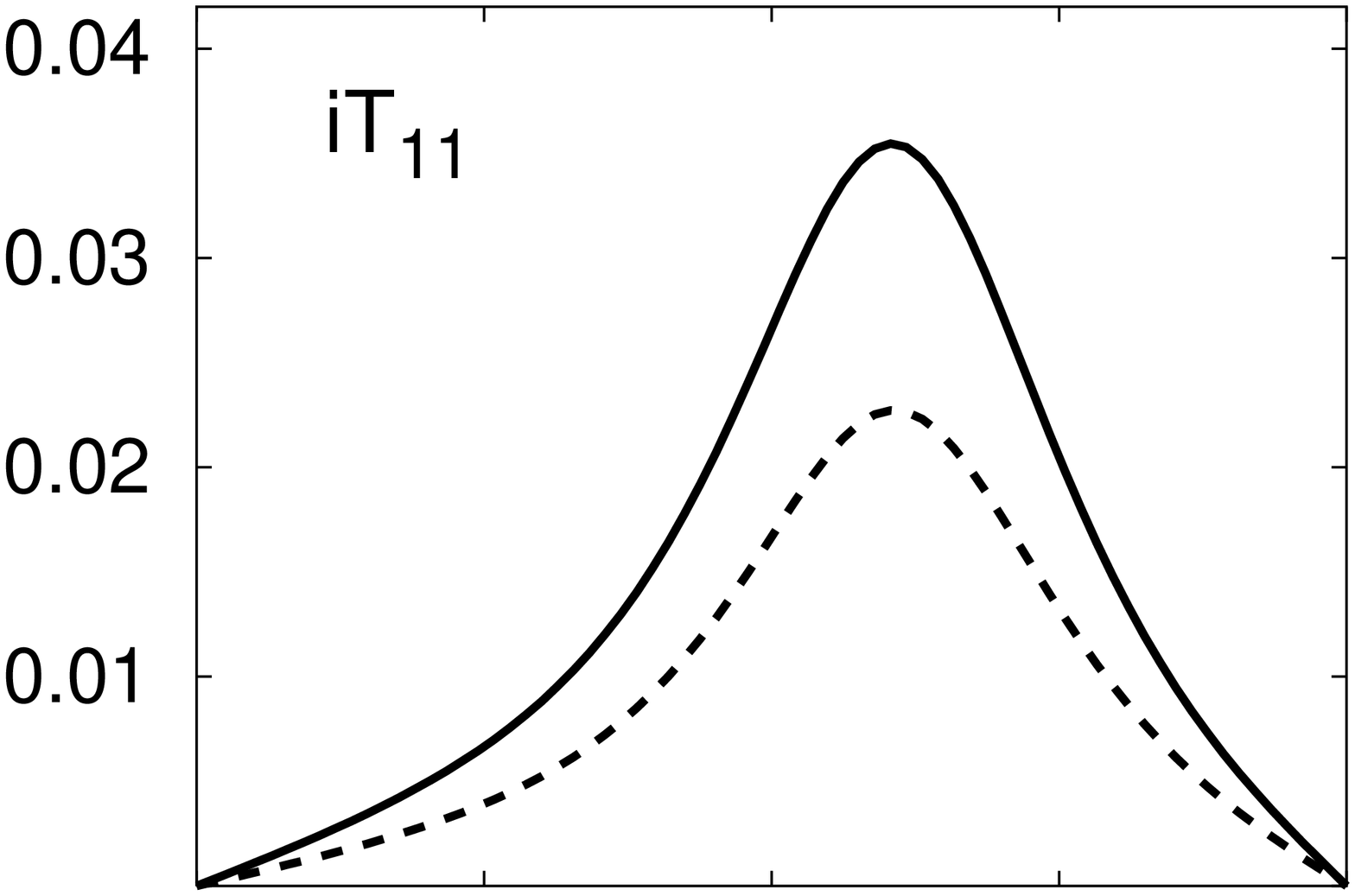,width=48mm,angle=-90}
}}
\vspace{-0.8cm}
\centerline{\hbox{
\psfig{figure=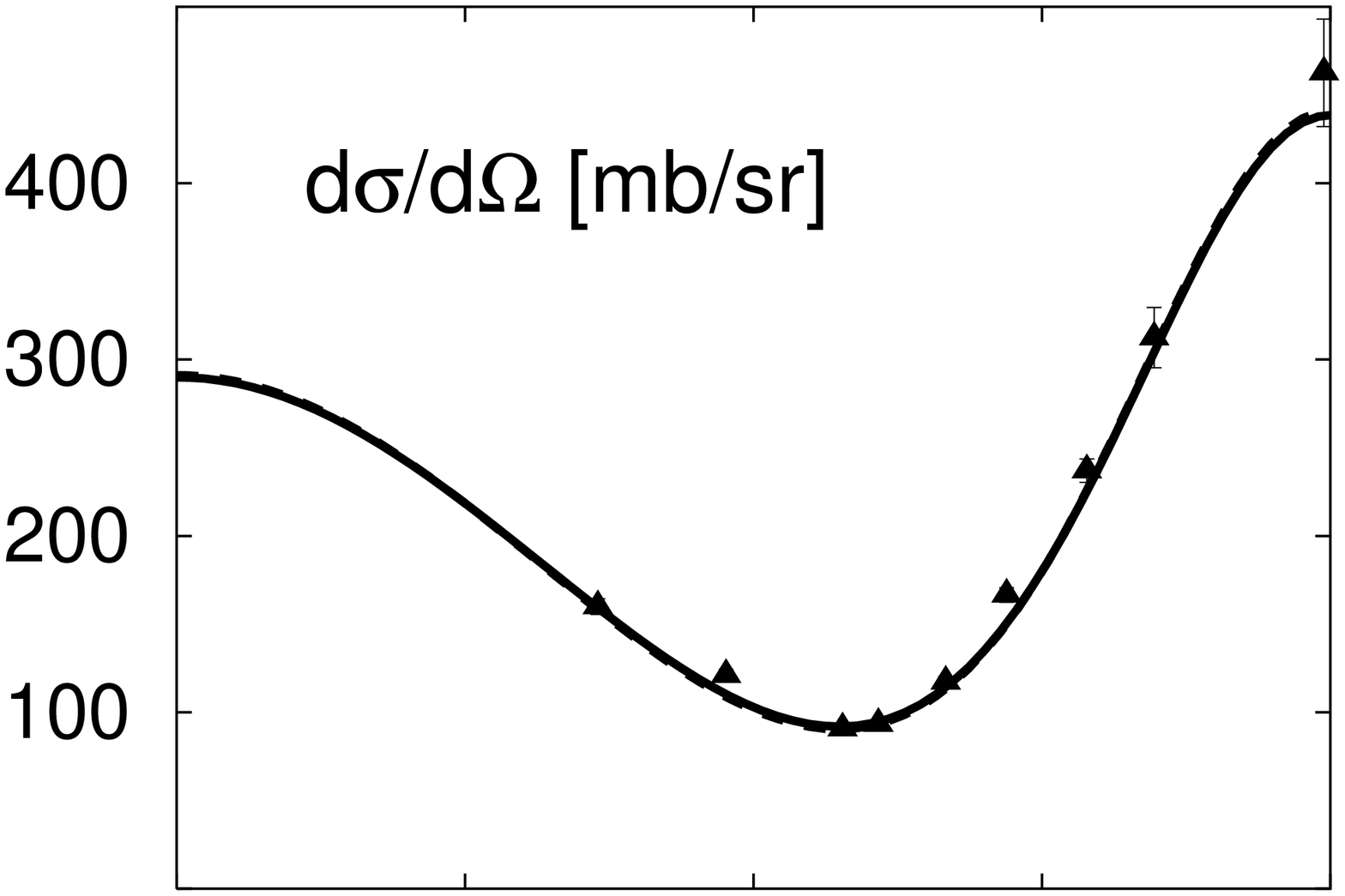,width=48mm,angle=-90}
\hspace{-1.0cm}
\psfig{figure=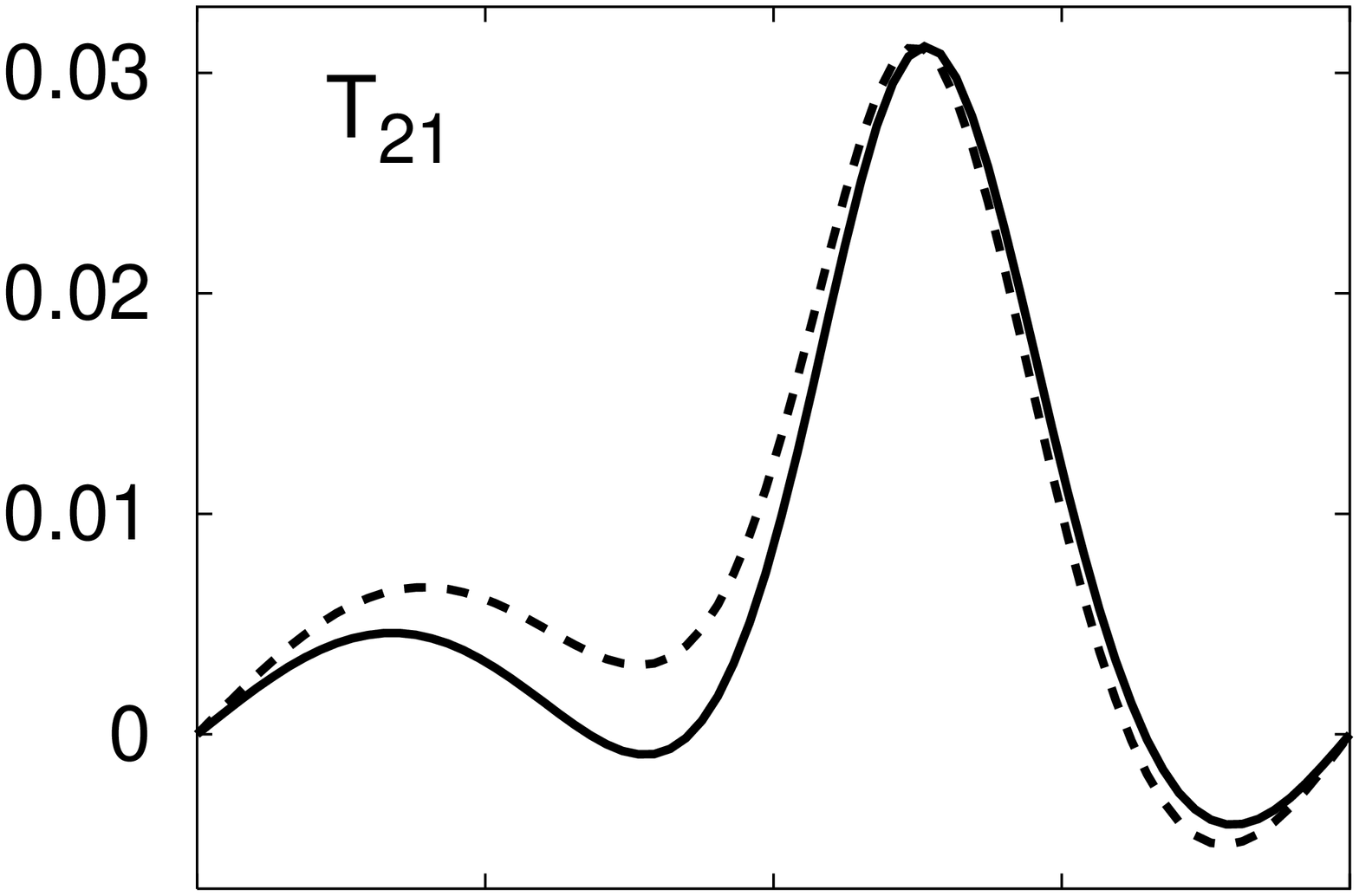,width=48mm,angle=-90}
}}
\vspace{-0.8cm}
\centerline{\hbox{
\psfig{figure=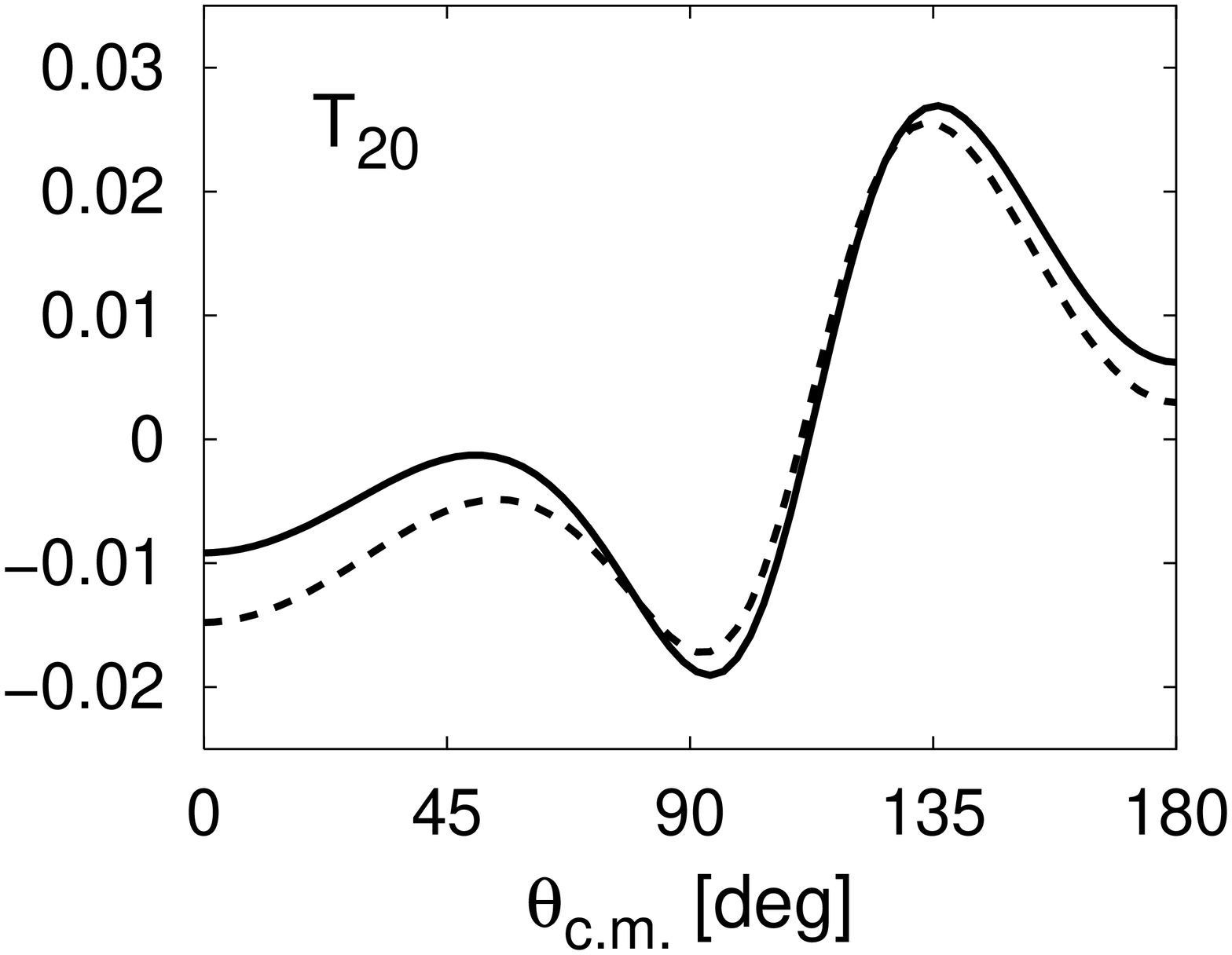,width=48mm,angle=-90}
\hspace{-1.0cm}
\psfig{figure=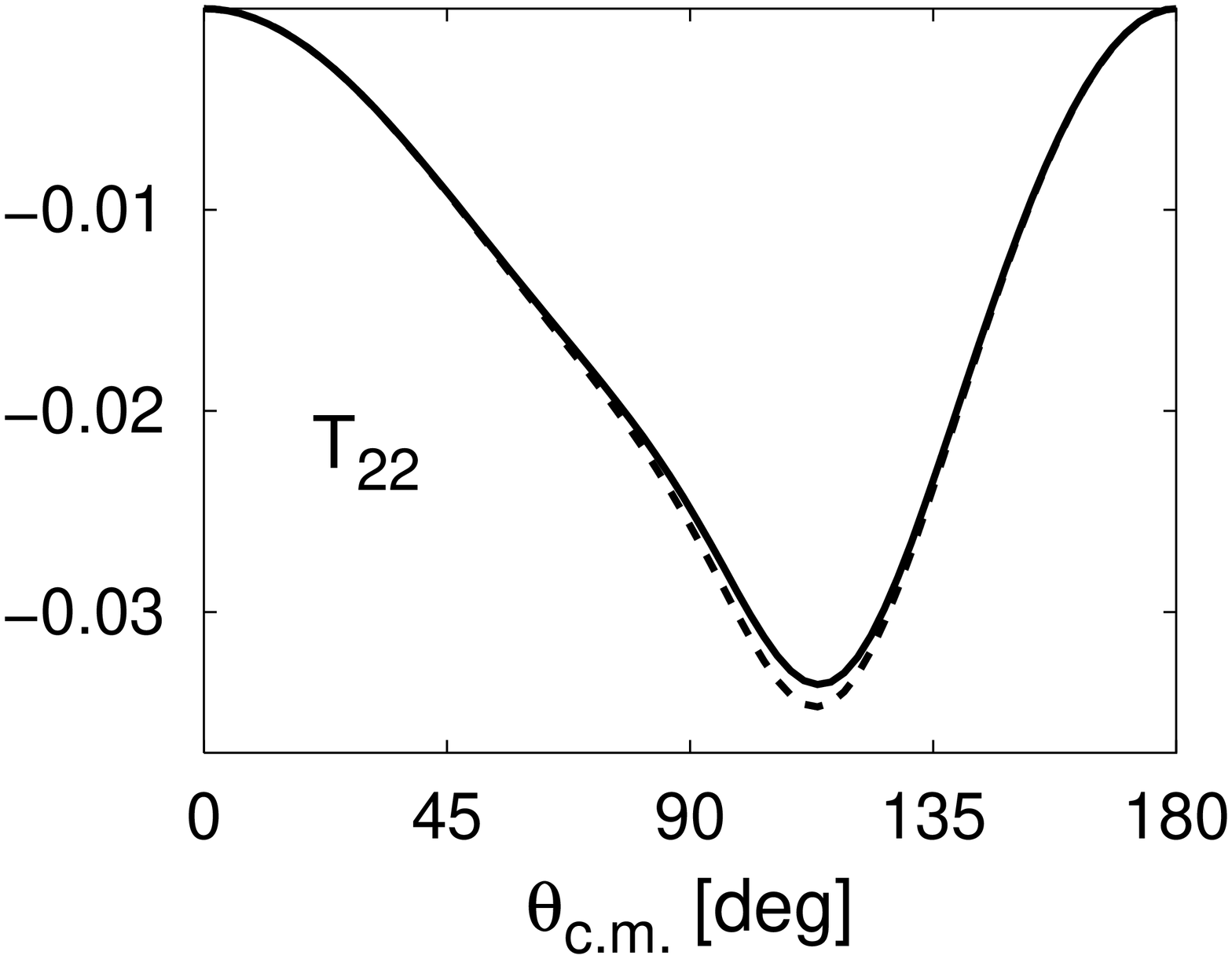,width=48mm,angle=-90}
}}

\vspace{11pt}
\caption[ ] {3$NF$ effects on the differential cross section and
analyzing powers, for $nd$ scattering at 3 MeV. 
The dashed lines are results with no $3NF$.  Solid line exhibits
$3NF$ effects as discussed in the text.}
\label{pan-pol}
\end{figure}

\noindent
integral equation
with the driving term calculated as follows
\begin{equation}
Z_{11}= \langle g_1| G_0 P |g_1\rangle +
\langle g_1| G_0 V^{3N}_1 G_0 |g_1\rangle \, .
\end{equation}
\noindent
The first contribution is the standard $2N$ driving term,
with $G_0$ and $P$ being the free Green's function
and the cyclic/anti-cyclic  permutator, respectively,
while the second takes into account the effects of the $3N$ force.
This procedure is consistent with
the formalism developed in Ref.\cite{CMS00}
to include irreducible pionic effects in the
$3N$ dynamics, and at the same time it corresponds to
the established perturbative procedure to include
$3NF$ effects in the separable AGS equation~\cite{OryuYamada}.

The results obtained are shown in Fig.~\ref{pan-pol},
for the differential cross-section, and the analyzing powers
$A_y$, $iT_{11}$, $T_{20}$, $T_{21}$, and $T_{22}$.
The panel exhibits the results obtained with the BBEST
potential, where a rank-4 expansion has been used for all
states with $j\le 2$, aside for the states 
$^1S_0$ (rank 5), $^3S_1$-$^3D_1$ (rank 6), and $^3P_2$-$^3F_2$ (rank 5).

The results without the $3NF$ are shown by the dashed lines,
and are practically indistinguishable from the corresponding
results for the Bonn {\em B} potential.
The solid curve considers the additional contribution
of the $3N$ force expressed in Eq.~(\ref{OPE-$3NF$}).
We have used the subtraction method of Eq.~(\ref{SUB-1}), 
with $t$ and $v$ given by the Bonn {\em B} interaction
while the parameter $c(Z)$ has been set to 0.73 for this energy.
Results obtained with the PEST/Paris potential are similar.

\begin{figure}
\centerline{\hbox{
\psfig{figure=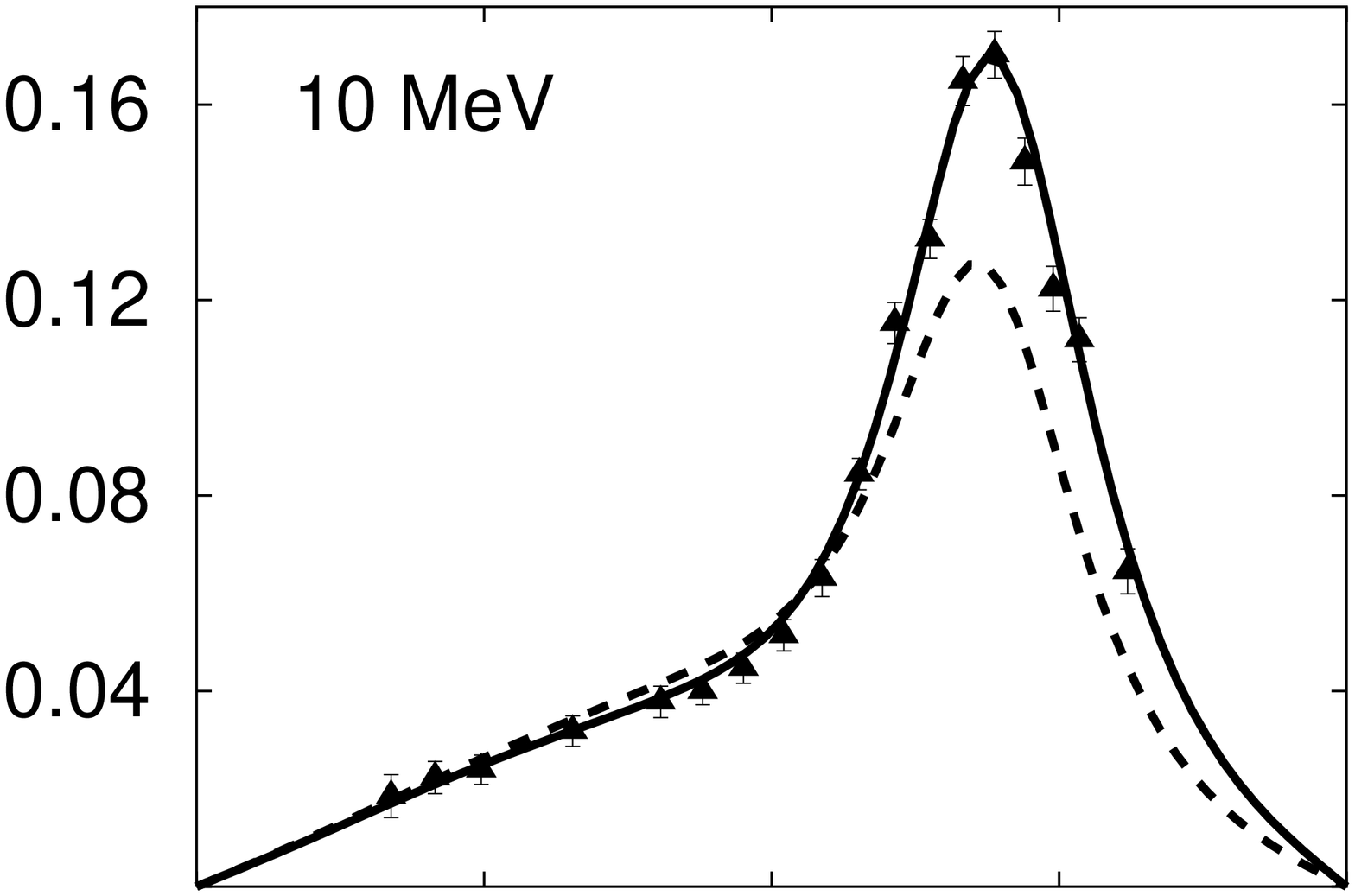,width=50mm,angle=-90}
}}
\vspace{-0.8cm}
\centerline{\hbox{
\psfig{figure=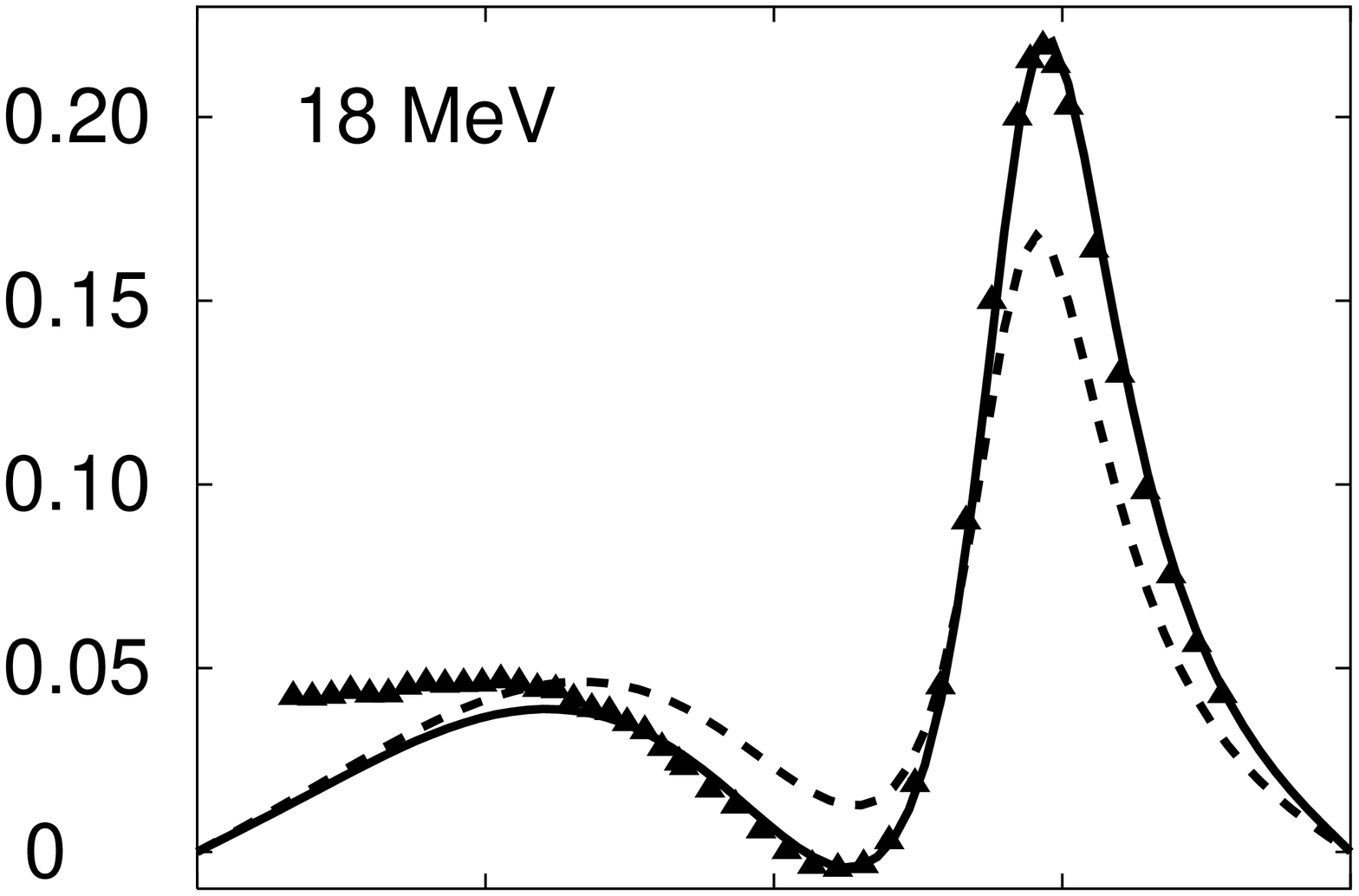,width=50mm,angle=-90}
}}
\vspace{-0.8cm}
\centerline{\hbox{
\psfig{figure=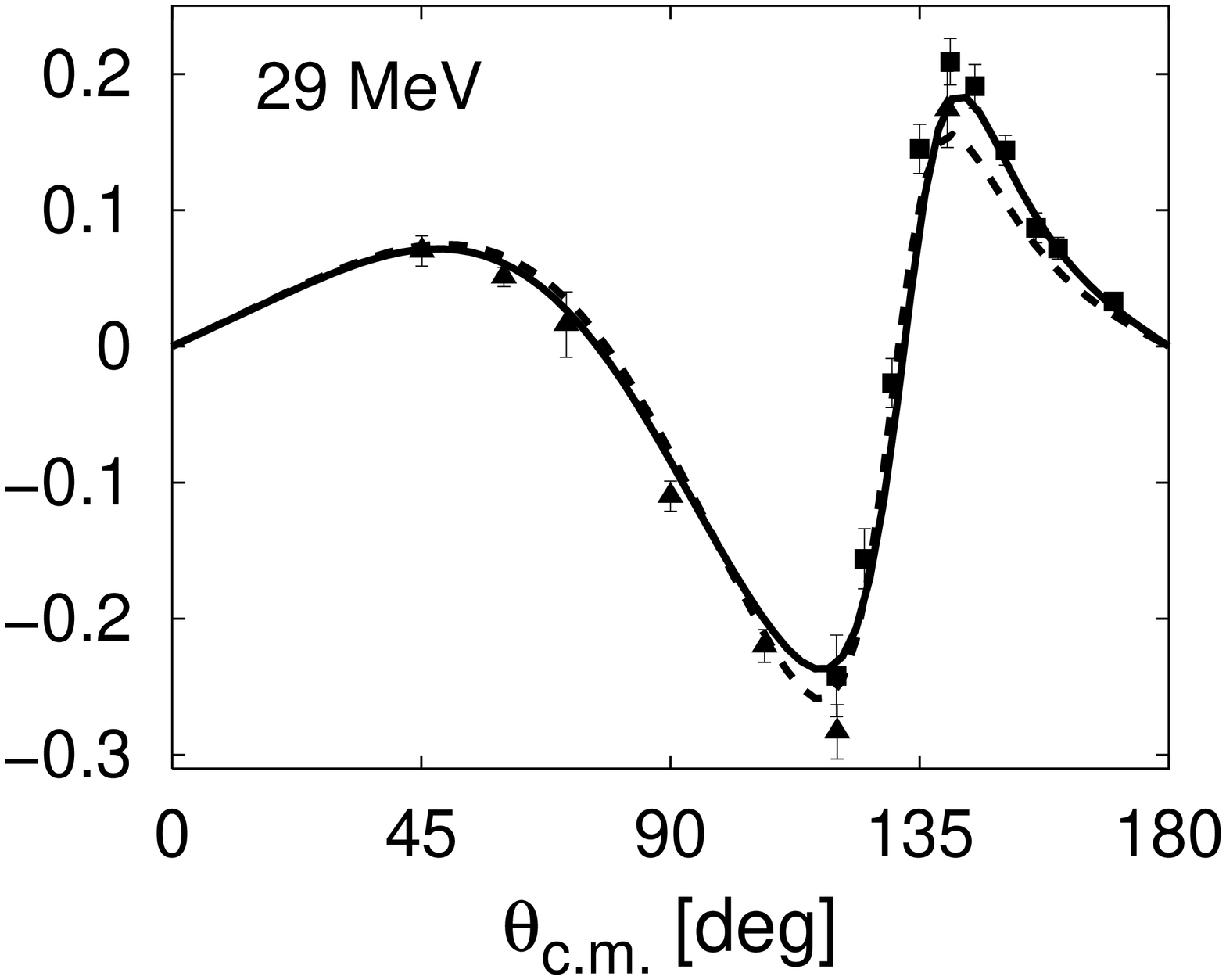,width=50mm,angle=-90}
}}
\vspace{11pt}

\caption[ ] {Same as in previous figure for $A_y$ at
10, 18, and 29.6 MeV.}
\label{pan-Ay}
\end{figure}

In Fig.~\ref{pan-Ay} we show how the puzzle evolves above 
the break-up  threshold, up to 30 MeV. In the top panel we 

\noindent
compare our theoretical calculations with the experimental 
data at 10 MeV~\cite{Tornow82}, obtaining the correct
reproduction for $A_y$ with the $3NF$ when $c(Z)=0.735$.
In the middle panel comparison is made with $pd$ data 
at 18 MeV~\cite{Sagara2001}. Comparison of $nd$ calculations with $pd$
data is somewhat questionable because of the perturbations
introduced by the Coulomb field. However it is known that, 
aside for the angles in forward direction, one of the main effects of 
the Coulomb field in $A_y$ can be approximately reproduced by
comparing the charged data with calculations performed 
at energies lowered of about 0.5-0.7 MeV. For this reason
the data are compared with calculations at 17.3 MeV. (However, care must be 
exercized in interpreting this fact as a Coulomb slow-down 
effect~\cite{Vlahovic99}) At this energy we obtain the reproduction of the 
observable with $c(Z)=0.773$. The botton panel shows the results obtained at
29.6 MeV, compared with the corresponding data at the same 
energy~\cite{Dobiash78} (triangles), and with $pd$ data at 30.2 MeV 
\cite{Johnston65} (squares). The solid line has been obtained with 
$c(Z)=0.81$. It is evident that as the energy increases,
the $c(Z)$ parameter shifts slowly towards one, as expected. 

We checked also how the triton binding energy is affected, and found
that here the effects are relatively small. With the highest possible rank,
{\it i.e.}, with a rank 5 representation in all states with  $j\le 2$, 
except the coupled states $^3S_1$-$^3D_1$ (rank 6), and $^3P_2$-$^3F_2$ 
(rank 7), the BBEST+$3NF$ result is $-8.137$ while the 
corresponding $2NF$ result is $-8.090$ (MeV).

We observe that some unavoidable approximations entered in this
study.  For instance, in obtaining Eq.~(\ref{OPE-$3NF$}) we have
ignored nucleonic recoil effects and have divided the subtracted
$t$-matrix by the pion mass, instead of $\omega_\pi$, to get simpler
expressions in partial waves.  In addition, there might be a possible 
$3NF$ contribution of shorter range for the exchange of a
$\rho$-meson; this term whould then counteract the one-pion-exchange
$3NF$, at least in the tensor part.  Other uncertainties are related to 
the perturbative treatment of the pion dynamics in the AGS 
equation~\cite{CMS00}. Finally, uncertainties about the
fully off-the-energy-shell extrapolation of the $NN$ $t$-matrix entering
in this $3NF$ forced us to introduce the effective parameter $c(Z)$.

As discussed in Refs.~\cite{CantonSchadow00,CMS00}, 
the $3NF$ contribution analyzed in
this study is just one class of irreducible diagrams generated by the
pion dynamics, and in a more complete analysis also other classes of
$3NF$ diagrams should be taken into account.  Forces of the type
TM~\cite{TM}, Brasil~\cite{BR}, Urbana~\cite{UA}, belong to another
class and they complement the $3NF$ whose effects have been calculated
here separately.  Some of these $3NF$'s give small corrections to
$A_y$, but not always in the right direction~\cite{Rosati96,Huber99}; 
their relevance,
however, appears to be greater elsewhere (e.g., the binding energy of
the triton).  There are also additional $3NF$ terms of shorter range
that might contribute, and in particular those obtained from
$\chi$PT~\cite{Huber99} appear very promising in providing an
additional correction to $A_y$, possibly in the right direction.

To summarize and conclude: we have evaluated here for the first time a
new ``pionic'' effect in the $3N$ system. The effect is a natural
consequence of a recently developed theory~\cite{Canton98,CMS00} for
the combined $\pi$-$3N$ dynamics in the $3N$ system, and has been
recast into a $3NF$ term of new structure by the
authors~\cite{CantonSchadow00}.  The underlying $3NF$ diagram
complements the extensively discussed $2\pi$-$3NF$ diagrams, and this
complementarity shows up in the way this force affects the $3N$
observables: while the $2\pi$-$3NF$ terms have a large contribution on
the $3N$ binding energy and little effects (in the considered energy range)
on the vector analyzing powers, 
we have shown here that this new $3NF$ term greatly modifies in
particular these two spin observables, and has the potential to
provide in full the solution of the $A_y$ puzzle.  Conversely, we
checked also that the same force produces smaller changes for the
triton binding energy. Since both effects are clearly needed for
describing the low-energy behaviour of the $3N$ system, it will be
important to investigate at this point what will be the effect of the
{\it combined} treatment of these two forces. 

\vspace{-15pt}

\acknowledgements
\vspace{-4mm}
This work is supported by the Italian MURST-PRIN Project ``Fisica
Teorica del Nucleo e dei Sistemi a Pi\`u Corpi".  W.~Sch. thanks INFN
and the University of Padova for hospitality and acknowledges support
from the Natural Science and Engineering Research Council of Canada.
L.~C. thanks TRIUMF for hospitality.

\end{document}